\RequirePackage{fix-cm}
\documentclass[smallextended]{svjour3}       
\smartqed  
\usepackage{amsmath}
\usepackage{bm}
\usepackage{ulem}

\usepackage{graphicx}
\begin{document}

\title{A novel application of probabilistic teleportation: $p$-Rabin qubit-oblivious transfer%
\thanks{Project supported by the National Key Research and Development Program of China (Grant No.2017YFB080200), National Natural Science Foundation of China (Grant No.61602378), and Natural Science Basic Research Plan in Shaanxi Province of China(Grant No.2016JQ6033)}
}


\author{Zhang MeiLing \and Li Jin \and Liu YuanHua \and Shi sha \and Zheng Dong \and Zheng QingJi \and Nie Min
}


\institute{M. Zhang, J. Li, Y. Liu, D. Zheng, Q. Zheng, M. Nie \at
              School of Communication and Information Engineering,
              XI'AN University of Posts \& Telecommunications, Xi'an 710121, China. \\
          S. Shi \at
          Engineering Research Center of Molecular and Neuro Imaging of Ministry of Education
          of China and School of Life Science and Technology, Xidian University, Xian 710071, China. \\
              \email{zhangmlwy@126.com}
}

\date{Received: date / Accepted: date}

\maketitle

\begin{abstract}
 All existing quantum oblivious transfer protocols are to realize the oblivious transfer of bit or bit-string. In this paper, $p$-Rabin quantum oblivious transfer of a qubit (abbr. $p$-Rabin qubit-OT) is achieved by using a probabilistic teleportation (abbr. PT) protocol. Here, this is the first time that the concept of qubit-OT is presented. As the PT protocol is able to transfer an (un)known pure state with a certain probability,  this feature makes the PT protocol well fit for Rabin OT.
 Furthermore, the PT protocol can be used for OT of a bit by encoding classical bit with two pre-agreed orthogonal states.
 Finally, security analysis shows that the PT protocol can be against participants attacks,
 and what's more, the discussion of relationship with no-go theorem demonstrates that the PT protocol is able to evade the no-go theorem.

\keywords{$p$-Rabin oblivious transfer \and probabilistic teleportation \and security analysis \and no-go theorem}
 \PACS{03.67.Dd   \and 03.67.Hk}
\end{abstract}

\section{Introduction}

Oblivious transfer(OT), firstly introduced by Rabin\cite{Rabin81}, is an important primitive in cryptography, in particular two-party computation. There are two major types of OT: Rabin OT (often referred to as all-or-nothing OT) and 1-out-of-2 OT. In Rabin OT, a sender (named Alice) wants to send a one-bit message $s$ to a receiver (named Bob) in such a way that Bob learns $s$ with probability $1/2$ at the end of the protocol.  Alice does not know whether Bob receives the $s$ or not, but Bob knows. In 1-out-of-2 OT, Alice sends two one-bit messages $m_0$ and $m_1$ to Bob,
and Bob can choose one of the messages but not knowing the other one,
while Alice doesn't know Bob's choice.
Later, these two OT protocols are shown to be equivalent in classical level \cite{Crepeau88}. \par

 As summarized in \cite{YHChou17}, realization methods of Quantum OT(QOT) can be roughly classified into $5$ types: \par
(1) PR non-locality box\cite{Buhrman06,ChouYH11,Gisin13}.\par
(2) Cr\'epeau's reduction \cite{YangL13,YangYG14,YangYG15_1,YangYG15_2,YangYG15_3} \par
(3) QBC-based QOT. Because of MLC no-go theorem\cite{Mayers97,LoChau97}, unconditional secure bit commitment is impossible within the scope of non-relativistic physics, which makes researchers try other ways to circumvent the no-go theorem. Later, based on relativistic, several unconditional secure BC protocols are proposed \cite{Kent99,Kent03,Kent12} and demonstrated experimentally \cite{Lunghi13,LiuY14}. So now here comes the question, could unconditional secure QBC lead to secure QOT?
  The analysis of He\cite{HeGP15} shows that the answer is not completely negative: some of the no-go proofs remain valid, while some other no-go proofs no longer work. As a result, this open question needs more attention.
\par
(4) bit-string QOT. In 2015,  based on quantum state computational distinguishability \cite{Kawachi12}, Souto et. al.\cite{Souto15_1,Souto15_2} presented a bit-string OT, which is pointed out unsuitable for reduction to 1-out-of-2 OT using Cr\'epeau's reduction\cite{Crepeau88} by He\cite{He2015} and Plesch et.al.\cite{Plesch17}. The main reason is that if the sender Alice cheats, the receiver Bob doesn't know that Alice also knows that Bob doesn't get the bit-string. Moreover, Plesch et.al.\cite{Plesch17} introduces an improved reduction that is appropriate for some flawed protocols such as \cite{Souto15_1} converting to 1-out-of-2 OT. \par
(5) Practical QOT\cite{Bennett92,Damgard05,LYB14,Roringues17}. The practical QOT protocols are based on some difficult problems, such as computational hard problems, technological limitations and so on.\par
As far as all-or-nothing QOT protocols are concerned, the first all-or-nothing QOT was proposed by  Cr\'epeau and Kilian in 1988\cite{CrepeauK88}.
 Damgard et al. proposed an efficient, non-interactive and secure all-or-nothing QOT in a bounded quantum storage model\cite{Damgard05}.
 To make Damgard＊s protocol tolerate the loss and error in both of quantum channel and measurements, the secret bit is covered by a correctable string in Li＊s protocol\cite{LYB14}.
 In 2015, an all-or-nothing bit-string QOT was presented, under the assumption of quantum hardness of state distinguishability and the constraint of performing at most few-qubit measurements\cite{Souto15_1}.
 In the same year, another all-or-nothing QOT of a bit is proposed with relatively few quantum resources for the receiver\cite{YangYG15_2}.
 Jo\~ao proposed an all-or-nothing bit-string QOT. Based on it, an all-or-nothing QOT of a bit is derived\cite{Roringues17}.
 The following two protocols utilize entangled states. An all-or-nothing QOT is proposed by using $n$ 4-qubit entangled states, and its security can be achieved if the $n$ is large enough\cite{He2006}. Under the assistance of an untrusted third party, Yang et al. proposed a secure all-or-nothing QOT \cite{YangYG14}.

All QOT protocols mentioned above are OT of a classical bit or a classical bit-string. In our knowledge, there is no related research about OT of a qubit yet. We find that the PT protocol\cite{LiW} fits for the definition of all-or-nothing OT very well. And the security analysis shows that the PT protocol can be indeed used to achieved  $p$-Rabin OT of a qubit.

 The rest of this paper is organized as follows.
 Section $2$ firstly introduces a typical PT protocol.
 Then its application in $p$-Rabin qubit-OT is illustrated.
 The security of the application is analyzed in Section $3$.
 Finally, a conclusion is drawn in Section $4$.
\section{ Probabilistic teleportation protocol and its application in $p$-Rabin qubit-oblivious transfer}
\subsection{a typical probabilistic teleportation protocol} \par
In\cite{LiW}, a protocol for PT of a qubit is presented using a partially entangled state. An (un)known qubit can be transmitted from a sender (Alice) to a receiver (Bob) via a quantum channel and LOCC. Suppose that Alice and Bob share a partially entangled pair
\begin{equation}
|\phi\rangle_{AB} = a|00\rangle_{AB} + b|11\rangle_{AB},  \ (|a|^2+|b|^2=1, |a|>|b|) \label{equ1}
\end{equation}
where Alice has particle $A$ and Bob has particle $B$.
 Alice has a qubit $C$ in pure state $|\zeta\rangle_C  = \alpha |0\rangle + \beta |1\rangle \ (|\alpha|^2+|\beta|^2=1)$ that will be teleported to Bob.
 Then the combined three particle system can be described as
 \begin{equation}
 |\phi\rangle_{CAB} = (\alpha a|000\rangle +\alpha b|011\rangle +\beta a|100\rangle +\beta b|111\rangle )_{CAB}
 \end{equation}
 Using the Bell state basis, $ |\phi\rangle_{CAB}$ can be rewrote as
  \begin{equation}
  \begin{aligned}
 |\phi\rangle_{CAB} = \frac{1}{\sqrt2} \Big[ & \ \ \ |\psi_1\rangle_{CA}(\alpha a|0\rangle + \beta  b|1\rangle))_{B}
                                                +   |\psi_2\rangle_{CA}(\alpha a|0\rangle - \beta  b|1\rangle))_{B} \\&
                                               + |\psi_3\rangle_{CA}(\beta  a|0\rangle + \alpha b|1\rangle))_{B}
                                               + |\psi_3\rangle_{CA}(\beta  a|0\rangle - \alpha b|1\rangle))_{B}   \Big]
 \end{aligned}
 \label{equ3}
 \end{equation}
where the four Bell states are
 \begin{equation}
  \begin{aligned}
 & |\psi_1\rangle = \frac{1}{\sqrt2} (|00\rangle + |11\rangle), \ \  |\psi_2\rangle = \frac{1}{\sqrt2} (|00\rangle - |11\rangle), \\&
   |\psi_3\rangle = \frac{1}{\sqrt2} (|01\rangle + |10\rangle), \ \  |\psi_4\rangle = \frac{1}{\sqrt2} (|01\rangle - |10\rangle).
 \end{aligned}
 \end{equation}
 The specific steps of the PT protocol are described as follows.
\newtheorem{Protocol}{Protocol}
\begin{Protocol}[A typical probabilistic teleportation protocol \cite{LiW}]
 \item {\textbf{Step 1}} Alice performs a joint Bell measurement(BM) on particles $C$ and $A$. The possible BM results, occurring probabilities and the corresponding collapsed states of particle $B$ are shown in Table~\ref{Table1}.
 \item {\textbf{Step 2}} Alice sends the result of BM to Bob.
 \item {\textbf{Step 3}} Bob first introduces an auxiliary qubit with the initial state $|\delta\rangle_m = |0\rangle_{m}$. If the BM result is $| \psi_i \rangle \ (i=1,2,3,4)$, an optimal 2-qubit unitary operator $U_i$ is applied on system $B,m$, taking the state $\frac{1}{\sqrt {Pr_i}}|\phi_i\rangle_B|0\rangle_{m}$ to $|\phi\delta_i\rangle_{Bm}$ given by
     \begin{equation}
     |\phi\delta_i\rangle_{Bm} =\Bigg\{
     \begin{aligned}
     & \frac{1}{\sqrt{2Pr_1}}\big[b(\alpha |0\rangle + \beta |1\rangle)_B |0\rangle_m  + \sqrt{1-2|b|^2}\alpha|0\rangle_B|1\rangle_m \big],  \  (i=1,2) \\&
       \frac{1}{\sqrt{2Pr_3}}\big[b(\alpha |0\rangle + \beta |1\rangle)_B |0\rangle_m  + \sqrt{1-2|b|^2}\beta|0\rangle_B|1\rangle_m \big],   \  (i=3,4).
     \end{aligned}
      \label{equ5}
     \end{equation}
     Then Bob measures particle $m$ using $Z$ basis. If the measurement result is $|0\rangle_m$, particle $B$ collapses to the teleported state $|\zeta\rangle_B =\alpha |0\rangle + \beta |1\rangle$. Therefore, Bob successfully recovers the teleported state with probability of $2|b|^2$.
\label{protocol1}
\end{Protocol}
\begin{table}[htbp]
 \caption{The BM results and  occurring probabilities and the corresponding collapsed states of particle $B$ }
 \centering
 \begin{tabular}{c | cccc}
   \hline
    BM result   & $| \psi_1 \rangle$
                & $| \psi_2 \rangle$
                & $| \psi_3 \rangle$
                & $| \psi_4 \rangle$ \\
   \hline
    $Pr_i$      & $\frac{1}{2} \big[ |\alpha a|^2 + |\beta b|^2 \big]$
                &  $\frac{1}{2} \big[ |\alpha a|^2 + |\beta b|^2 \big]$
                & $\frac{1}{2} \big[ |\beta a|^2 + |\alpha b|^2 \big]$
                &  $\frac{1}{2} \big[ |\beta a|^2 + |\alpha b|^2 \big]$ \\
   \hline
    $|\phi_i\rangle_B$   & $\frac{1}{\sqrt{2Pr_1}}(\alpha a|0\rangle + \beta  b|1\rangle)$
                       & $\frac{1}{\sqrt{2Pr_2}}(\alpha a|0\rangle - \beta  b|1\rangle)$
                       & $\frac{1}{\sqrt{2Pr_3}}(\beta  a|0\rangle + \alpha b|1\rangle)$
                       & $\frac{1}{\sqrt{2Pr_4}}(\beta  a|0\rangle - \alpha b|1\rangle)$ \\
   \hline
   $U_i$      & $U_1$
              & $U_2$
              & $U_3$
              & $U_4$ \\
   \hline
   $Pr_{suc}$         & $\frac{1}{2}|b|^2$
                      & $\frac{1}{2}|b|^2$
                      & $\frac{1}{2}|b|^2$
                      & $\frac{1}{2}|b|^2$  \\
   \hline
    $Pr_{fail}$       & $\frac{1}{2}|\alpha|^2(1-2|b|^2)$
                      & $\frac{1}{2}|\alpha|^2(1-2|b|^2)$
                      & $\frac{1}{2}|\beta|^2(1-2|b|^2)$
                      & $\frac{1}{2}|\beta|^2(1-2|b|^2)$  \\
   \hline
 \end{tabular}
  \label{Table1}
 \end{table}
The unitary transformations $U_i \ (i=1,2,3,4)$ in Table~\ref{Table1}  are given as
\begin{equation}
\begin{aligned}
& U_1= \left[
\begin{matrix}
             A(a,b)        &    \bm{0}   \\
             \bm{0}      &     \sigma_z
\end{matrix}
\right],
& U_2= \left[
\begin{matrix}
              A(a,b)        &    \bm{0}   \\
             \bm{0}       &     -\sigma_z
\end{matrix}
\right]   \\
& U_3= \left[
\begin{matrix}
           \bm{0}        &    \sigma_z   \\
            A(a,b)         &    \bm{0}
\end{matrix}
\right],
& U_4= \left[
\begin{matrix}
           \bm{0}        &    -\sigma_z   \\
            A(a,b)         &    \bm{0}
\end{matrix}
\right] \\
\end{aligned}
\end{equation}
where $\bm{0}$ is the $2 \times 2$ zero matrix, $\sigma_z$ is the phase-flip operator, and $A(a,b)$ is the $2 \times 2$ matrix relative to the parameters $a$ and $b$ of Equ.~\ref{equ1}. $\sigma_z$ and $A(a,b)$ are expressed as
  \begin{equation}
 \sigma_z = \left[
 \begin{matrix}
 1  &  0 \\
 0  &  -1
 \end{matrix}
 \right]  \ \
 A(a,b) = \left[
 \begin{matrix}
     \frac{b}{a}         &   \sqrt{ 1 - \frac{|b|^2}{|a|^2}} \\
      \sqrt{ 1 - \frac{|b|^2}{|a|^2}}    &    -\frac{b}{a}
 \end{matrix}
 \right]
 \end{equation}
The four $U_i(i=1,2,3,4)$ satisfy the following property
\begin{equation}
U_1 = V_1 U_2 = V_2 U_3 = V_3 U_4,
\end{equation}
where $V_1, V_2$ and $V_3$ are given by
  \begin{equation}
 V_1 = \left[
 \begin{matrix}
 I  & \  \bm{0} \\
 \bm{0}  & \  -I
 \end{matrix}
 \right]   \ \
 V_2 = \left[
 \begin{matrix}
 \bm{0}  & \  I \\
 I  & \  \bm{0}
 \end{matrix}
 \right]      \ \
 V_3 = \left[
 \begin{matrix}
 \bm{0}  & \  I \\
-I       &  \ \bm{0}
 \end{matrix}
  \right];
   \ \text{and} \ \
   I = \left[
   \begin{matrix}
   1  & \ \  0 \\
  0       &  \ \ 1
   \end{matrix}
    \right]
 \end{equation}

 \newtheorem{Lema}{Lema}
\begin{Lema}
For the receiver Bob, Protocol.~\ref{protocol1} provides an optimal strategy to extract the teleported state and the maximal probability of successfully extracting the teleported qubit is $2|b|^2$ \cite{LYH}.
\end{Lema}
\subsection{An novel application of the Protocol.~\ref{protocol1}}
Another description of Protocol.~\ref{protocol1} to fit for the $p$-Rabin qubit-OT is given as follows.
 \textbf{Qubit to transfer:} $|\zeta\rangle_C  = \alpha |0\rangle + \beta |1\rangle \ (|\alpha|^2+|\beta|^2=1)$. \par
 \textbf{Quantum channel:} $|\phi\rangle_{AB} = a|00\rangle + b|11\rangle,  \ (|a|^2+|b|^2=1, |a|>|b|)$. The coefficients $a$ and $b$ only affect the probability of Bob's receiving the transferred qubit, so there is no need to keep them secret. \par
 \textbf{Step 1} Alice and Bob securely share a partially entangled pairs $|\phi\rangle_{AB}$. \par
 \textbf{Step 2} Using $|\phi\rangle_{AB}$ as quantum channel,  Alice teleports $|\zeta\rangle_C$ to Bob by implementing  Protocol.~\ref{protocol1} . \par
 \textbf{Step 3} Bob will recover  $|\zeta\rangle_B$ in particle $B$ with probability of $2|b|^2$. \par
The above description illustrates that the Protocol.~\ref{protocol1} fits the definition of OT very well.
Moreover,  $p$-Rabin bit-OT can also be achieved by simply encoding classical bits $0$ and $1$ with two orthogonal qubits such as $|+\rangle$ and $|-\rangle$.
However, the orthogonal basis \{$|0\rangle$, $|1\rangle$\} cannot be used to encode classical bits $0$ and $1$.
 According to Table~\ref{Table1}, if the $|\zeta\rangle_C$ is $|0\rangle$, i.e. $\alpha=1$ and $\beta=0$ ,  the state received by Bob
is $|0\rangle$ when Alice's BM result is $|\psi_1\rangle$ or $|\psi_2\rangle$, or
   $|1\rangle$ when Alice's BM result is $|\psi_3\rangle$ or $|\psi_4\rangle$.
Therefore, Bob can recover the transferred qubit with probability $1$ by applying Pauli operator $I$ or $X$ according to Alice's BM results.
Similarly, if the $|\zeta\rangle_C$ is $|1\rangle$, i.e. $\alpha=0$ and $\beta=1$, Bob can also recover the transferred qubit with probability $1$. \par
For the clarity of description, the term $p$-Rabin qubit-OT protocol is used to replace Protocol.~\ref{protocol1} in the  following sections.

\section{Security analysis of the $p$-Rabin qubit-OT protocol}
In this section, we analyze the security of the $p$-Rabin qubit-OT protocol (i.e. Protocol.~\ref{protocol1}). With reference to the requirements of $p$-Rabin bit-OT \cite{Crepeau88}, the $p$-Rabin qubit-OT should satisfy the following four properties(the first express the correctness while the last three assure the security of the protocol):
\begin{description}
\item{\bf{Soundness:}} If Alice and Bob are both honest, then Bob will obtain the right message with probability $p$. While Bob knows whether he gets the right message or not, Alice is oblivious of the fact. \par
\item{\bf{Concealingness:}} If Alice is honest, Bob cannot learn the right message before the opening phase. \par
\item{\bf{Probabilistic transfer:}} After opening phase, Bob cannot learn the right message with probability greater than $p$.\par
\item{\bf{Oblivious:}} If Bob is honest, after the opening phase Alice must not know with certainty whether Bob received the right message or not.
\end{description} \par

\centerline{ \bf{(A) Soundness}}\par
It's clear that the $p$-Rabin qubit-OT protocol fulfills the soundness criterion while the probability $p=2|b|^2$. As the coefficients $a$ and $b$ are related to the probability, it's important to make sure the correctness of the shared quantum channel $|\phi\rangle_{AB}$. If the quantum channel and classical channel between Alice and Bob are both authenticated, the securely sharing of $|\phi\rangle_{AB}$ can be simply realized as follows. \par
(1) In order to pre-share $n$ entangled states of $|\phi\rangle_{AB}$, Alice prepares $n+m$  $|\phi\rangle_{AB}$ and separates them into two sequences $S_A$ and $S_B$. Both of them include $n+m$ particles. \par
(2) Alice randomly inserts $k$ decoy states to $S_B$ to get $S_B'$ and sends $S_B'$ to Bob. \par
(3) After Bob receives the $S_B'$, Alice publishes the locations and encoding basis of the decoy states. Then Bob checks the decoy states to prevent attacks from the quantum channel. If there are no attacks from the channel, Bob is able to recover $S_B$. \par
(4) Bob randomly chooses $m$ locations and asks Alice to send the corresponding qubis in $S_A$ to him. Then Bob measures the corresponding $m$ $|\phi\rangle_{AB}$ with the orthogonal basis of ($|\eta_1\rangle,|\eta_2\rangle,|\eta_3\rangle,|\eta_4\rangle$), where $|\eta_1\rangle=|\phi\rangle_{AB}$, $|\eta_2\rangle=b|00\rangle-a|11\rangle$, $|\eta_3\rangle=a|01\rangle+b|11\rangle$, $|\eta_4\rangle=b|01\rangle-a|11\rangle$. All the measurement results should be $|\phi\rangle_{AB}$, or else attacks exist in the channel or Alice may cheat. As a result, the above process of sharing quantum channel repeats again. \par
It's a more complicated case if Alice and Bob are strangers, i.e. there is no authenticated channels between them. An quantum entanglement establishment protocol is proposed to solve the case\cite{Hwang2016}.

\centerline{ \bf{(B) Concealingness}}\par
In the $p$-Rabin qubit-OT protocol, the opening phase corresponds to Step 2 in Protocol.~\ref{protocol1}, i.e. Alice's sending the BM result to Bob.
Before receiving the BM result from Alice, Bob's state $|\phi\rangle_B$ is in a mixed state $\{Pr_i, |\phi_i\rangle_B \} (i=1,2,3,4) $ shown in Table~\ref{Table1}. Bob's state can also be represented by the density matrix
\begin{equation}
\rho_B = \sum_{i=1}^4  Pr_i |\phi_i\rangle_B \langle \phi_i|
=  \left[
 \begin{matrix}
 |a|^2  &  0 \\
 0      &  |b|^2
 \end{matrix}
 \right]
\end{equation}
which gives no information about the right state. \par

\centerline{ \bf{(C) Probabilistic transfer}}\par
After opening phase, Bob learns the exact state of his particle $B$. For example, if Alice notices Bob that her measurement result is $|\psi_2\rangle$, then Bob knows that his particle is in the state $|\phi_2\rangle_B = \frac{1}{\sqrt{2Pr_2}}(\alpha a|0\rangle - \beta  b|1\rangle)$. According to \textbf{Lema 1}, for Bob, the maximal probability of successfully extracting the right qubit is $p=2|b|^2$.   \par
Note that if the OT of the same qubit repeats $n$ times, the probability of successfully extracting the qubit is given by
$ 1-(1-2|b|^2)^n$. The probability increases with the increase of $n$, so there is no reason for Bob to ask Alice to retransfer the qubit.
If there exists some external factors that affect the correctness of the received qubit, such as loss and noise occurring in the channel, tamper from attackers, some techniques should be used to overcome these factors.
 For example,  error correction  can be used against the loss and noise occurring in the channel and
 the tamper can be detected by using decoy technique. \par
\centerline{ \bf{(D) Oblivious}}\par
Now we prove that the $p$-Rabin qubit-OT protocol is oblivious:  Alice does not know whether Bob received the right qubit or not.
Note that Alice's attacks shouldn't lead to the result that Bob will not be able to know of receiving the right state with certainty, as it violates
 the original intention of OT. \par
If Alice launches attacks after Bell state measurement, because of Bob's performing only local operations and measurements, Alice has no way of knowing Bob's measurement result of his auxiliary qubit.
But there is still a case that needs to be taken into account: in the opening phase, if Alice tells Bob a fake BM result,
the probability of successful OT would decrease or not.  Suppose that Alice's BM result is $|\psi_i\rangle$ and the state of qubit $B$ collapses to $|\phi_i\rangle_B$,
but she tells a fake one $|\psi_j\rangle (j \neq i)$ to Bob. After receiving the fake message, Bob applies $U_j$ on particle $B$, i.e. $U_j|\phi_i\rangle_B|0\rangle_m$.
According to Equ.$8$ and $9$, we can derive a matrix $W$, the element $W_{ji}(j,i=1,2,3,4)$ of which is used to describe the relationship between $U_j$ and $U_i$, i.e. $U_j=W_{ji}U_i$.
  \begin{equation}
 W = \left[
 \begin{matrix}  \ \
 I       & \ \   V_1     & \ \   V_2       & \ \   V_3   \\
 V_1     & \ \    I      & \ \   V_1V_2    & \ \   V_1V_3 \\
 V_2     & \ \  V_2V_1   & \ \   I         & \ \   V_2V_3 \\
 -V_3    & \ \  -V_3V_1  & \ \   -V_3V_2   & \ \   I
 \end{matrix}
 \right]
 \end{equation}
 Therefore, $U_j |\phi_i\rangle_B|0\rangle_m = W_{ji}U_i |\phi_i\rangle_B|0\rangle_m$, then Bob gets $W_{ji}|\phi \delta_i\rangle_{Bm}$, where $|\phi \delta_i\rangle_{Bm}$ is shown in Equ.5.
 For example, $i=2, j=3$, i.e. $W_{ji}=W_{32}=V_2V_1$, then Bob gets the following state
 \begin{equation}
 \begin{aligned}
    W_{32}|\phi \delta_2\rangle_{Bm} & = V_2V_1
      \frac{1}{\sqrt{2Pr_1}}\big[b(\alpha |0\rangle + \beta  |1\rangle)_B |0\rangle_m  + \sqrt{1-2|b|^2}\alpha|0\rangle_B|1\rangle_m \big]   \\
 & =  \frac{1}{\sqrt{2Pr_1}}\big[b(-\beta |0\rangle + \alpha |1\rangle)_B |0\rangle_m  + \sqrt{1-2|b|^2}\alpha|1\rangle_B|1\rangle_m \big].
 \end{aligned}
 \end{equation}
 Thus, when Bob's measurement result of particle $m$ is $|0\rangle$, qubit $B$ collapses to  $-\beta |0\rangle + \alpha |1\rangle$.
 Bob believes that he recovers a right quantum message but actually not, which violates the original intention of OT.
\par
Next, we consider the following two cases that the attacks are to launch before the Bell measurement.
 Before the Bell measurement, Alice can launch two types of attacks that may be able to know whether Bob would receive the right qubit at the end of the protocol.   \par
The first possible attack is the unitary operation attack in which Alice applies an unitary operator $U_A$ on particle $A$ before Bell measurement. By this attack, Alice hopes that she can control the probability of Bob's successful extraction of the right qubit. \par
Let $U_A = k_1 I + k_2 X + k_3 Z + k_4 iY$, where $I, X, Z, iY$ are Pauli operators\cite{Nielsen} which can connect the four Bell states. The detail relations are shown in Table~\ref{Table2}.
  \begin{table}[htbp]
 \caption{The relations between Pauli matrices and Bell states}
 \centering
 \begin{tabular}{c | c | c | c | c}
   \hline
        & $|\psi_1\rangle$ & $|\psi_2\rangle$  &  $|\psi_3\rangle$  &    $|\psi_4\rangle$ \\  \hline
   $I$  & $|\psi_1\rangle$ & $|\psi_2\rangle$  &  $|\psi_3\rangle$  &    $|\psi_4\rangle$ \\  \hline
   $X$  & $|\psi_3\rangle$ & $|\psi_4\rangle$  &  $|\psi_1\rangle$  &    $|\psi_2\rangle$ \\  \hline
   $Z$  & $|\psi_2\rangle$ & $|\psi_1\rangle$  &  $-|\psi_4\rangle$  &    $-|\psi_3\rangle$ \\  \hline
  $iY$  & $|\psi_4\rangle$ & $|\psi_3\rangle$  &  $-|\psi_2\rangle$  &    $-|\psi_1\rangle$ \\  \hline
 \end{tabular}
   \label{Table2}
 \end{table}
 \par
 According to Equ.~\ref{equ3},  $|\phi\rangle_{ABC} = \sum_{i=1}^{4} |\psi_i\rangle_{CA} |\phi_i\rangle_B$.
By applying $U_A$ on particle $A$ of $|\phi\rangle_{ABC}$, the state of the tripartite system $ABC$ becomes
\begin{eqnarray}
|\Lambda\rangle & = & U_A |\phi\rangle_{ABC} =\sum_{i=1}^{4} (U_A |\psi_i\rangle_{CA}) |\phi_i\rangle_B \nonumber \\
                & = & \ \  (k_1|\psi_1\rangle +  k_2|\psi_3\rangle  + k_3|\psi_2\rangle  + k_4i|\psi_4\rangle )_{CA} |\phi_1\rangle_B \nonumber \\
                &   & + (k_1|\psi_2\rangle +  k_2|\psi_4\rangle  + k_3|\psi_1\rangle  + k_4i|\psi_3\rangle )_{CA} |\phi_2\rangle_B \nonumber  \\
                &   & + (k_1|\psi_3\rangle +  k_2|\psi_1\rangle  - k_3|\psi_4\rangle  - k_4i|\psi_2\rangle )_{CA} |\phi_3\rangle_B \nonumber  \\
                &   & + (k_1|\psi_4\rangle +  k_2|\psi_2\rangle  - k_3|\psi_3\rangle  - k_4i|\psi_1\rangle )_{CA} |\phi_4\rangle_B \label{equ13}
\end{eqnarray}
According to Equ.~\ref{equ13}, suppose that Alice's BM result is $|\psi_1\rangle$, particle $B$ collapses to
\begin{eqnarray}
|\phi'\rangle_B & = &  \frac{1}{\sqrt{2}}\big[(k_1+k_3)\alpha + (k_2 + k_4i)\beta  \big] a |0\rangle
                       +\frac{1}{\sqrt{2}}\big[(k_1-k_3)\beta  + (k_2 - k_4i)\alpha \big] b |1\rangle  \nonumber \\
                & \overset{\triangle}{=} & \delta_1 a |0\rangle + \delta_2 b |1\rangle
\end{eqnarray}
After receiving Alice's BM result, Bob performs $U_1$ on $|\phi'\rangle_B |0\rangle_m$ and get $\delta_1 |0\rangle + \delta_2  |1\rangle$ with probability $\frac{1}{2}|b|^2$. Bob believes that he receives a right quantum message but actually not, which violates the original intention of OT. \par
The other possible attack is entanglement-measure attack in which Alice entangles an auxiliary qubit $E$ with particle $A$.
\begin{eqnarray}
|\Gamma\rangle & = & CNOT_{AE}|\phi\rangle_{ABC}|0\rangle_E \nonumber \\
               & = & (\alpha a|0000\rangle +\alpha b|0111\rangle +\beta a|1000\rangle +\beta b|1111\rangle )_{CABE} \nonumber \\
               & = &  \frac{1}{\sqrt2} \Big[|\psi_1\rangle_{CA}(\alpha a|00\rangle + \beta  b|11\rangle))_{BE}
                                                +   |\psi_2\rangle_{CA}(\alpha a|00\rangle - \beta  b|11\rangle))_{BE}  \nonumber \\
               &   &  \ \ \ \ + |\psi_3\rangle_{CA}(\beta  a|00\rangle + \alpha b|1\rangle))_{BE}
                                               + |\psi_3\rangle_{CA}(\beta  a|00\rangle - \alpha b|11\rangle))_{BE}   \Big] \label{equ11}
\end{eqnarray}
Here, $CNOT_{AE}$ is the controlled NOT gate, where $A$ is the control qubit and $E$ is the target qubit.
Without loss of generality, suppose that Alice's BM result is $|\psi_1\rangle$, particle $B$ and particle $E$ collapses to an entangled state
$ |\phi\rangle_{BE} = \alpha a|00\rangle_{BE} + \beta  b|11\rangle_{BE}$. Bob applies $U_1$ on particles $B$ and $m$, then the state of tripartite system $BEm$ becomes
  \begin{equation}
 |\phi_{BEm}\rangle =\frac{1}{\sqrt{2}} \big[b(\alpha|00\rangle +\beta|11\rangle)_{BE}|0\rangle_m
                                             + \alpha\sqrt{1-2|b|^2} |00\rangle_{BE}|1\rangle_m \big].
 \end{equation} \par
 If the result of Bob's measurement on particle $m$ is $|0\rangle_m$, the subsystem of particles $B$ and $E$ collapses to another entangled state
 $ |\phi\rangle_{BE} = \alpha |00\rangle_{BE} + \beta  |11\rangle_{BE}$. If Alice makes a measurement on particle $E$, Bob's state randomly collapses to corresponding $|0\rangle_B$ or $|1\rangle_B$, which also violates the original intention of OT. According to measurement result on particle $E$, Alice cannot judge whether Bob receives the right message or not .\par

 \centerline{ \bf{(E) relationship with no-go theorem}}\par
Here we discuss the relationship between the $p$-Rabin qubit-OT protocol and no-go theorem. On one hand, the MLC no-go theorem\cite{Mayers97} provides a  strategy to cheat the protocols such as BB84. In the $p$-Rabin qubit-OT protocol, the quantum channel is securely pre-shared. If Alice's target is to know whether Bob gets the right qubit, such attack that has been discussed in $(C)$ will fail. On the other hand, Lo's no-go theorem\cite{LoChau97} provides a strategy to cheat the protocols that satisfy definition 1. Corresponding to the related variable, the $i$ includes the qubit $C$ and a BM result, meanwhile, the $j$ is nothing in the $p$-Rabin qubit-OT protocol. Thus, the $p$-Rabin qubit-OT protocol cannot be viewed as the protocol in definition 1. The reasons are as follows (compare the items  one by one with those in definition 1).\par
 (I) Alice inputs a qubit and a BM result, while Bob inputs nothing; \par
 (II) Bob learns $f(i,j)$ with probability $50\%$ instead of certainty; \par
 (III) Alice knows that the $j$ is nothing; \par
 (IV) Bob can learn part of the $i$, i.e. the BM result.
\newtheorem{Definition}{Definition}
\begin{Definition}\textbf{Ideal one-sided two-party secure computation} \par
(I)Alice inputs $i$ and Bob inputs $j$, then \par
(II) Bob learns $f(i,j)$ unambiguously.\par
(III) Alice learns nothing about $j$ and $f(i,j)$. \par
(IV) Bob learns nothing about $i$ more than what logically follows from the values of $j$ and $f(i,j)$.
\end{Definition}

\section{Conclusion}
In this paper, by utilizing the property of PT, A typical PT \cite{LiW} is used to realize OT of a qubit from Alice to Bob. The $p$-Rabin qubit-OT protocol can easily lead to a QOT of a classical bit by encoding classical bits 0 and 1 with two orthogonal qubits,
so a new idea is given to realize QOT. This research may open a new perspective of OT. \par
To analyze the security of the $p$-Rabin qubit-OT protocol, we prove that the protocol satisfies the security requirements of a Rabin OT. Because the kernel part of the protocol is PT, the concealingness and probabilistic transfer is clear.
To learn the information whether Bob receives the teleported state or not, Alice may have three methods, one is to launch after BM and the other two are to launch before BM:(1)In the opening phase, Alice tells Bob a fake BM result in order to reduce the probability of Bob's recovering the right state; (2)Applying a unitary transformation on particle $A$ for the purpose that she can control the probability of Bob's receiving the right state; (3) Introducing an auxiliary qubit which is used to load and illustrate the information, so the auxiliary qubit must correlate to the pre-shared $|\phi\rangle_{AB}$. In all, the analytic results show that the $p$-Rabin qubit-OT protocol is unconditionally secure against any cheating strategy. Finally, the discussion of the relationship with the no-go theorem indicates the $p$-Rabin qubit-OT protocol is able to evade the no-go theorem.




\end{document}